\documentclass{elsart}

\usepackage{graphicx,epsfig,pstricks,rotating}
\usepackage{amsmath}
\usepackage{amssymb}

\newcommand{\beq}{\begin{equation}}
\newcommand{\eeq}{\end{equation}}
\newcommand{\bdm}{\begin{displaymath}}
\newcommand{\edm}{\end{displaymath}}
\newcommand{\beqa}{\begin{eqnarray}}
\newcommand{\eeqa}{\end{eqnarray}}

\begin{document}

\begin{frontmatter}

\title{ Influence of corruption on economic growth rate and 
foreign investments
}

\author[label1,label2,label3]{Boris~Podobnik},
\author[label3]{Jia~Shao}, 
\author[label2]{Djuro~Njavro}, 
\author[label3,label4]{Plamen~Ch.~Ivanov}, 
\author[label3]{H. Eugene Stanley} 

\address[label1]{Department of Physics, Faculty of Civil Engineering,
 University of Rijeka, Rijeka, Croatia}

\address[label2]{Zagreb School of Economics and Management,
 Zagreb, Croatia} 
 
\address[label3]{Center for Polymer Studies and Department of
    Physics, Boston University, Boston, MA 02215}

\address[label4]{Institute of Solid State Physics, 
Bulgarian Academy of Sciences, Sofia, Bulgaria}    

\bigskip

\begin{abstract} 

In order to investigate whether government regulations against
corruption can affect the economic growth of a country, 
we analyze the dependence between Gross Domestic Product 
(GDP) per capita growth rates and
changes in the Corruption Perceptions Index (CPI). 
For the period 1999-2004
on average for all countries in the world, we find that an
 increase of CPI by one unit 
leads to an increase of the annual GDP per capita by
1.7 \%. By regressing only European transition countries, we 
find that $\Delta$CPI = 1  generates increase of the annual 
GDP per capita by 2.4 \%. 
We also analyze the relation between foreign direct investments received by
different countries and CPI, and we find a statistically significant power-law
functional dependence between foreign direct investment per capita and
the country corruption level measured by the CPI. 
We introduce a new measure to quantify the relative corruption between countries
based on their respective wealth as measured by GDP per capita.

\end{abstract}

\end{frontmatter}
Corruption, defined as abuse of public power for private benefit, is a
global phenomenon that affects almost all aspects of social and economic
life.  Examples of corruption include the sale of government property by
public officials, bribery, embezzlement of public funds, patronage and 
nepotism. 
 The World Bank estimates that over 1000 billion US dollars annually
are lost due to corruption, representing 5\% of the world GDP.  The
African Union estimates that due to corruption, the African continent loses
25\% of GDP.

Previous studies have mainly reported a negative association between
corruption level and country wealth \cite{Svensson,Mauro,TD,Jia}, i.e., on
average richer countries are less corrupt. There is ongoing debate  concerning 
the relation 
 between corruption and  economic growth \cite{Lambd}. 
 Some earlier studies
suggested that corruption may even help the most efficient firms bypass
bureaucratic obstacles and rigid laws \cite{Leff}, while recent 
 papers  do not find a significant  negative  association 
 between  growth and corruption  \cite{Svensson,Mauro}.  The majority of
studies have found an insignificant negative association between
the corruption level and foreign investments \cite{Mauro,Wheeler,Wei}, without 
reporting a specific functional dependence.

In order to find
a quantitative relation between corruption level and economic factors such as 
GDP growth rate and foreign direct investments, we analyze the 
Corruption Perceptions Index (CPI) \cite{CPI}  
introduced by Transparency International, a global civil organization
supported by government agencies, developmental organizations,
foundations, public institutions, the private sector, and
individuals. The CPI is a composite index 
ranging from 0 to 10, where 0 denotes the highest level of corrupt and 
10 corresponds to the
lowest corruption level. 
For GDP per capita we use 
annual nominal GDP per capita in current prices in
US dollars \cite{GDPcurr}, and GDP per capita in
 constant dollars \cite{FDI}. 
 
The CPI is an absolute measure of corruption which does not depend on 
country wealth.  
However, besides in absolute terms of corruption level
countries may be also compared in relative terms where 
corruption level is compared 
depending on the countries' wealth as measured by the GDP per capita.  

In Table I, we show the first ten least corrupt countries as ranked by 
Transparency International according to the CPI values obtained in 2006
as well as some other countries.
Besides some
Western European countries, among the least corrupt ten countries are
New Zealand, Singapore, and Australia.  Chile and Botswana are the least
corrupt countries in South America and Africa, 
whereas Singapore is the least corrupt
Asian country. Table I provides information about corruption levels
throughout the World in absolute terms, where each country, whether
rich or poor, is given only its CPI value.

In the modern economy, globalization leads to economic competition 
and comparison 
between countries, 
so we compare the corruption levels for different groups of 
countries in the world.
Normalizing the CPI value for year 2006 on the population in
 each country  \cite{pop}, 
we find a normalized CPI value for the world to be 3.7, 
for the countries in Europe 
we find 5.4, for Asia and Latin America we find 3.3, and for Africa 2.7.

In an earlier study some of us have reported
a power-law functional dependence between GDP per capita and CPI for all 
countries in the world \cite{Jia}: 
\begin{equation}
{\rm CPI} = N ~({\rm GDP}_{pc})^{\mu}
\end{equation}
with scaling exponent $\mu \approx 0.23$ [see Fig.~1]. This functional 
dependence spans multiple 
scales of wealth and remains stable over different time periods.
The positive value of exponent $\mu$ indicates that richer countries are
less corrupt. 
This power-law
dependence provides information about the expected level of corruption for a
given level of country wealth --- e.g., a country above (or below) the fitting
line is less (or more) corrupt than expected for its level of wealth. We
may say that for a country above the fitting line the level of corruption 
is less than the expected level for the given country wealth.

\begin{table}
\caption{Rank of countries (left colomn) by Transparency International for year 2006 with CPI values (right colomn) for each country.}
\smallskip
\begin{tabular}{ccc}
\hline
\hline
1  &Finland,~Iceland,~New Zealand &{\bf 9.6  } \\
4  &Denmark &{\bf 9.5 }\\
5 & Singapore&{\bf  9.4 } \\
6  &Sweden& {\bf 9.2 } \\
7  &Switzerland&{\bf  9.1 }  \\
8  &Norway&{\bf  8.8  } \\
9  &Australia,~Netherlands&{\bf  8.7 } \\
11  &United Kingdom & {\bf 8.6 } \\
16  &Germany& {\bf 8.0 } \\
17  &Japan& {\bf 7.6 } \\
18  &France,~Ireland&{\bf  7.4 } \\
20  &Belgium,~Chile,~USA &{\bf 7.3 }   \\
37  &Botswana &{\bf 5.6} \\
40  & Italy &{\bf 5.0} \\
70
&China, India, Mexico, Brazil, Ghana, Egypt,  Peru, S.Arabia,
 Senegal&       {\bf 3.3} \\
121 &   Russia& {\bf 2.5} \\
\hline
\hline
\end{tabular}
\label{table1}
\end{table}

This previous finding indicates that in order to compare the corruption
level between two countries, countries may be compared not
only in terms of absolute CPI values but also in terms of relative country wealth.  
To that account, we introduce a new measure of relative corruption which we call 
{\it Honesty~per~Dollar} (HpD): 
\begin{equation}
{\rm HpD} = \ln({\rm CPI}) - \mu \ln({\rm GDP_{pc}}) - c,  
 \end{equation}
equal to the difference between actual CPI and the value expected from
the power-law fitting line. 

\centerline{\epsfig{file=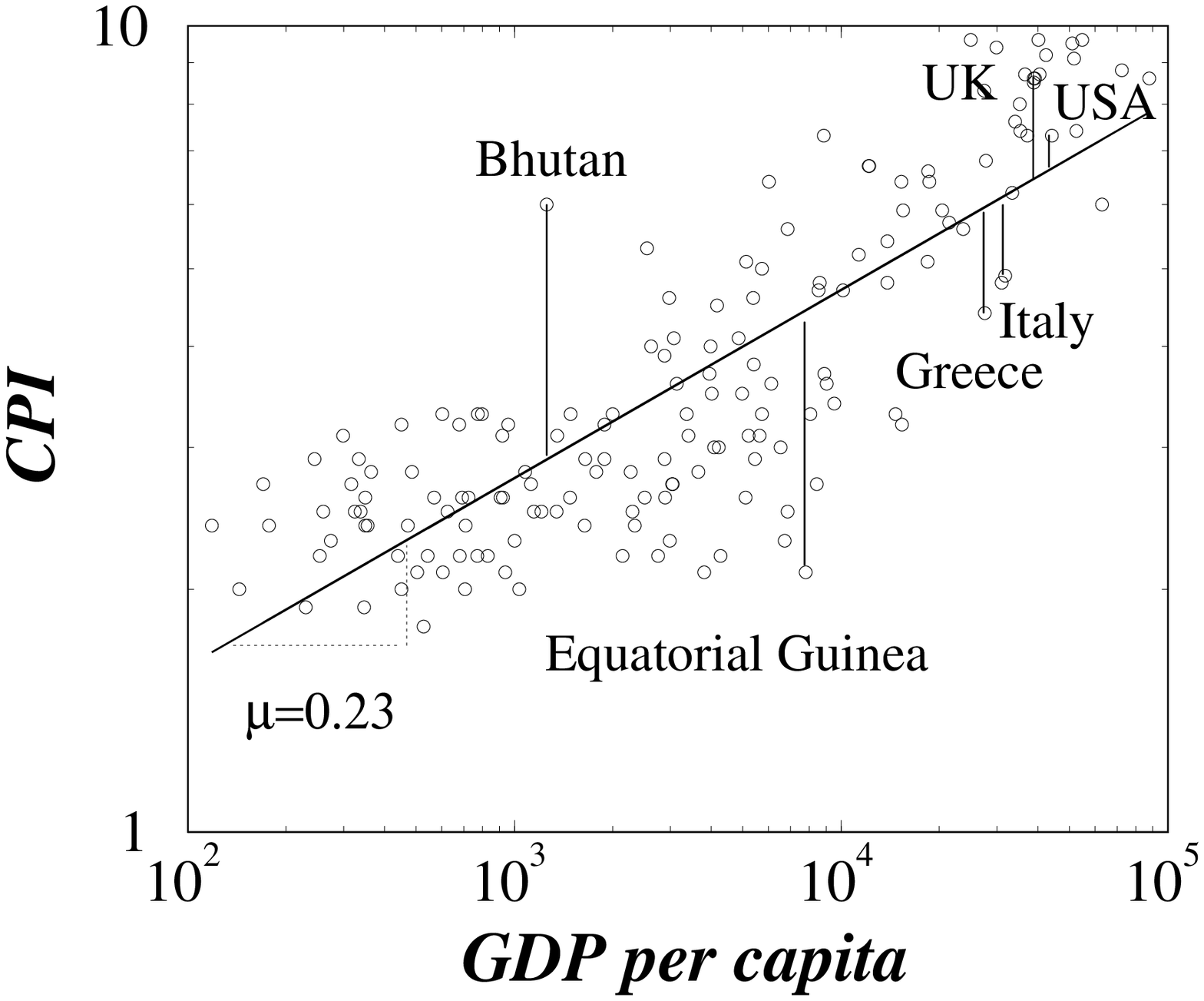,width=9cm}}
\label{fig.1}
{\footnotesize\textbf{Figure 1:} Corruption level measured by Corruption
Perceptions Index (CPI) versus country wealth measured by GDP per capita
calculated for 2006. We find the
functional dependence can be fit by a power law $0.56 ~(GDP_{pc})^{0.23}$
 with positive
exponent. The power law fit in log-log plot has an obvious statistical
explanation, representing the expected level of CPI for a country with
given GDP per capita. The countries that are above the line are less
corrupt than expected. We define a new index we call Honesty per Dollar
(HpD) to measure relative performance of a country when CPI and GDP per
capita are simultaneously considered.
 Besides the USA, UK, Greece, and Italy, we show 
 the countries with the extreme HpD values, Bhutan and Equatorial Guinea
 (oil exporter).
}

We assume that all countries, with similar GDP per capita and   
 laying on the power-law fitting line in Fig.~1, have comparable levels of
corruption when $({\rm HpD} =0)$. Generally, the larger value for HpD, the better
the performance of a country.  For 2006 based on regression obtained for the entire
world, we calculate the values of the index for some countries:
 HpD(UK) = $0.29$,    HpD(USA) = $0.1$, 
 HpD(Italy) = $-0.23$,   
 HpD(Greece) = $-0.3$. 
The negative values of HpD indices for Italy and Greece, indicate that these two
countries are relatively more corrupt than expected for their 
corresponding level of wealth (GDP per capita).
 
One of the  reasons for a country to reduce corruption is to
attract more foreign investments, and thus to additionally increase the GDP. 
 This is because corruption generally
increases start-up costs for new businesses. If  investors can choose
between two countries with different levels of corruption, they may
choose not to start their business in a more corrupt country since the
profit in that country will be reduced. In previous study  we have analyzed
how the corruption level relates to foreign direct investments received
by different countries from the United States ~\cite{Jia}.  For each continent we
have found that the functional dependence between the U.S. direct
investments per capita, {\it I}, and the corruption levels across
countries exhibits scale-invariant behavior characterized by a power law
$CPI\sim {\it I}^{\lambda}$. Since $\lambda > 0$ for each continent,
less corrupt countries have received on average more U.S. investment per
capita.

\centerline{\epsfig{file=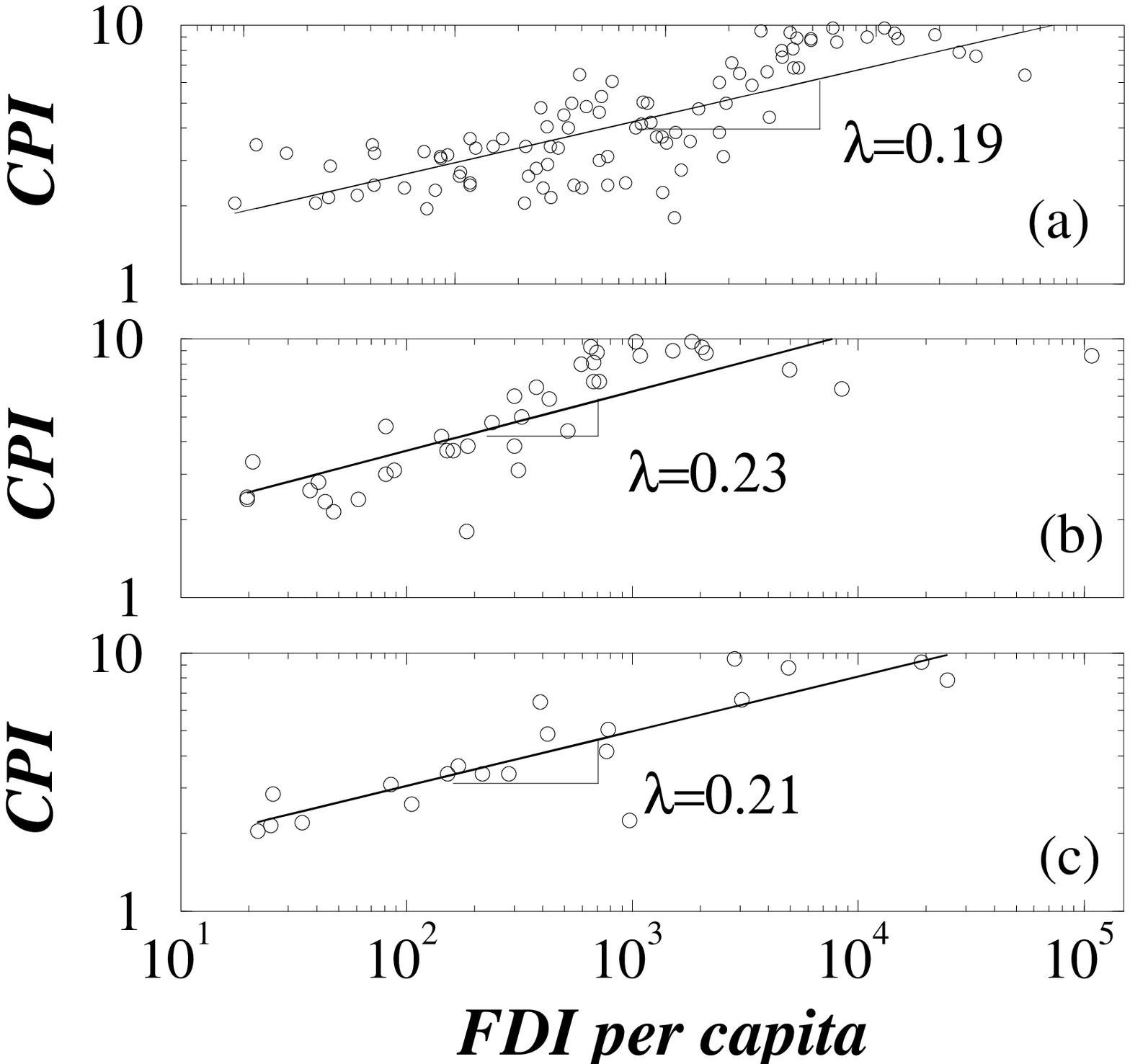,width=9.cm}}
\label{fig.2}
{\footnotesize\textbf{Figure 2:} Less corrupt countries receive more
foreign investments. For the period 1999-2004, 
we  show average foreign direct investments (FDI) per capita, 
denoted by $I$,  received
by (a) World, (b) European, and (c) Asian countries 
 from all foreign countries versus
corruption level measured by CPI.  We find the
statistically significant  power-law dependence between $I$ and $CPI$, 
$CPI\sim {\it I}^{\lambda}$ with scaling exponents:
  for the World $\lambda = 0.19~ (\Delta = 0.016)$, 
  Europe $\lambda = 0.23~ (\Delta = 0.029)$, 
  Asia $\lambda = 0.21 ~(\Delta = 0.029)$.
  In brackets we put the standard errors of the exponents.  
  In the study we exclude Indonesia   
 as a  country with total  negative  value for FDI and Cameroon.  
 }

Here, for each country in the world we analyze the foreign direct investments 
(FDI) received  from all foreign countries, not only from the
US. For each country we sum up the foreign direct  investments  
 over the period  1999-2004, and we calculate the 
 average FDI per year per capita.
 In Fig 2  we show that the functional dependence between
the average foreign direct investment per capita, I, and the corruption level
measured by CPI exhibits power-law behavior ${\rm CPI}\sim {\it
I}^{\lambda}$ with a statistically
significant scaling exponent $\lambda = 0.19$ and a standard error
$\Delta = 0.016$ \cite{tratio}. 
 As for the case
of the foreign direct investments originating from the US, we find that
less corrupt countries on average receive more foreign investments per
capita than more corrupt countries. We repeat the whole analysis but
this time for different continents. Again we obtain the power-law
 dependence  ${\rm CPI}\sim {\it I}^{\lambda}$ with scaling exponents:
  for Europe $\lambda = 0.23~ (\Delta = 0.029)$, Asia $\lambda = 0.21 ~(\Delta = 0.029)$,  
Latin America $\lambda = 0.23 ~ (\Delta = 0.085)$ and Africa  
$\lambda = 0.18 ~(\Delta = 0.059)$. 
In the parenthesis we put the standard errors from which we 
conclude that for each
 continent the  power-law exponent is statistically
significant at the $5 \%$ level. Note that the scaling exponent we
obtain for Europe is larger than  the scaling exponent $\lambda =0.14$ 
obtained for
the US foreign direct investments in Ref.~\cite{Jia}.

We investigate the relation between change in
 CPI and economic growth as measured by 
growth in the GDP per capita. 
For the period 1999-2004 and  world countries ranked by Transparency
International, we run regression fit between the 
change in the logarithm of the GDP per capita 
 in constant dollars as dependent variable and
the change in CPI for this period as the explanatory variable. 
In Fig.~3(a) we show GDP per capita  growth rates 
 versus  change in CPI 
 that can be fit by a linear regression with 
slope $\tau \approx 0.09$. We find that an increase in 
CPI by one unit leads on average to a 
$1.7\%$ increase in GDP per capita.

We perform the same analysis for 39 European countries ranked 
 by Transparency International
for the period 1999 to 2004 and we obtain
statistically insignificant  dependence  between 
 GDP per capita growth rate and  difference of CPI 
 (exponent $\tau = 0.036$ and 
 standard error $\Delta = 0.042$). Then we repeat the same analysis 
for 21 European developing  (transition) countries.  In Fig 3(b) for the
period 1999-2004 we show the GDP per capita growth rate in constant
dollars vs.  difference of CPI.  We find a  functional
dependence that can be approximated by a straight line, 
where the exponent
$0.12$ (standard error $ \Delta = 0.049$) is statistically significant 
at the $5 \%$ level. 
From the exponent obtained for 5 years period, the increase
of  CPI by one unit is followed by additional annual increase
of GDP per capita of approximately $2.4\%$.  Plot of the GDP
per capita growth rate in constant dollars vs.  difference in CPI with
similar statistically significant exponent $ \tau = 0.11$ with error
$\Delta = 0.044$, we find by analyzing all new EU members [see Fig.~3(b)].

\centerline{\epsfig{file=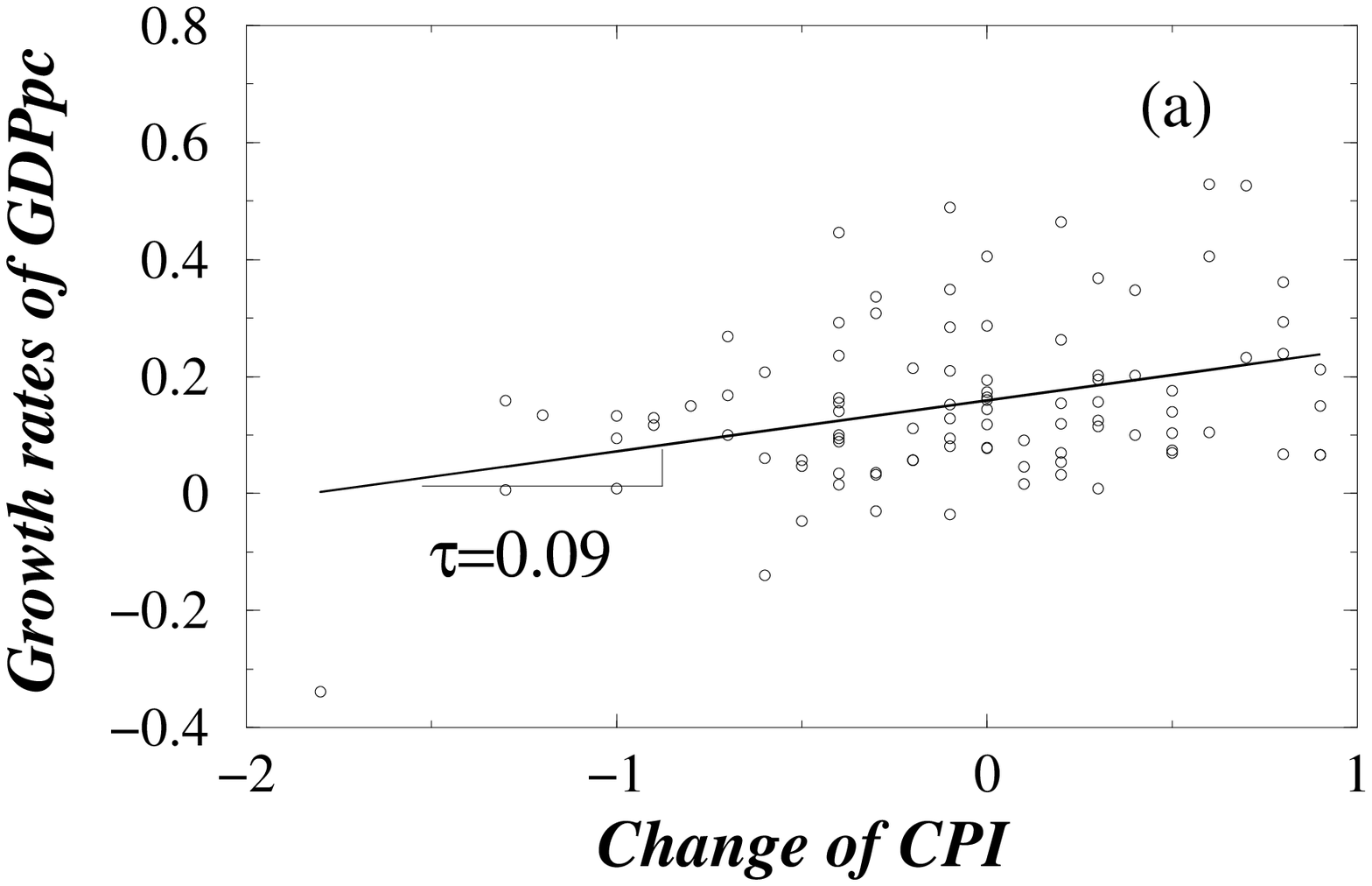,width=9cm}}
\vspace{10.mm} 
 \centerline{\epsfig{file=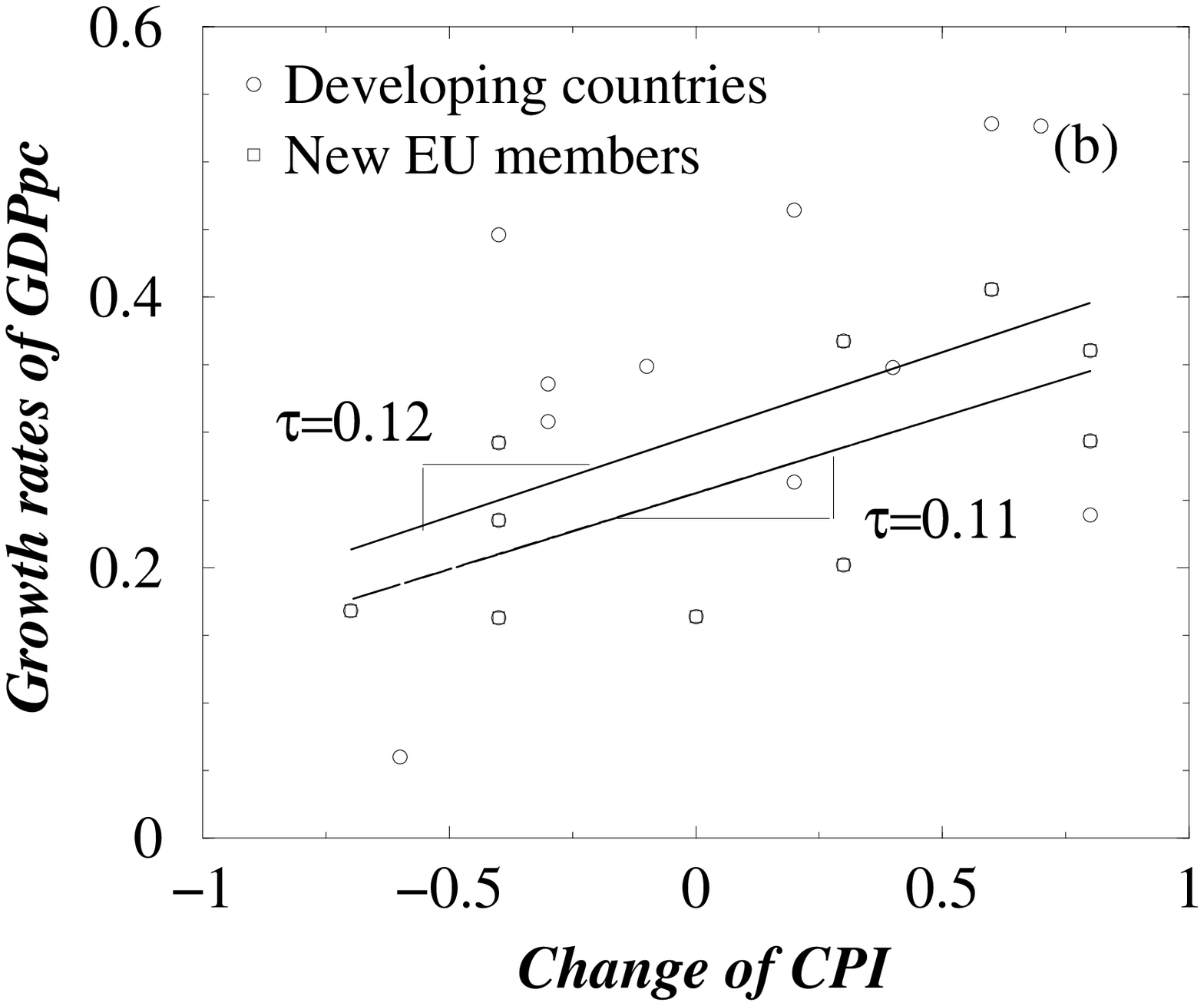,width=9cm}}
\label{fig.3}
{\footnotesize\textbf{Figure 3} Countries improving more corruption
level generates larger GDP per capita growth rate.  For the period
1999-2004, we plot growth rate of GDP per capita in
constant dollars versus difference of CPI.  We analyze 
(a) World countries (except Belgium and Uruguay) and (b) 
21 European transition  countries.  For each case we find a  functional
dependence that can be approximated by a straight line.  
For case (a)  by using linear regression we obtain   
exponent $\tau = 0.09$ (five years period) with standard error 
 $\Delta = 0.024$. 
For case (b),  we obtain  
exponent $\tau = 0.12$ (five years period) with  
 $\Delta = 0.049$.  Thus, for (b) we find   that -- on
yearly basis --- increase of  CPI by one is followed on average by increase
of GDP per capita equal to $\approx$ 2.4\%. Separately, for ten new EU 
members we obtain that the functional dependence between GDP per capita
growth rates and change of CPI  can be fit by linear
regression with statistically significant
 exponent $\tau = 0.11$  and standard error  $\Delta = 0.044$. 
Note that if Belgium and Uruguay (outliers) are included in (a),    
 the estimated exponent in this regression is $0.052$,  
where  $\Delta=0.022$. 
 }

In summary, we have observed a statistically significant power-law functional 
dependence between CPI 
and foreign direct investment per capita. 
This power-law dependence spans broad range of scales in 
 foreign direct investment (from hundreds to tens of thousands of dollars).  
We also find a statistically significant 
dependence between changes in CPI and GDP per capita
 growth rate, indicating that reducing the 
corruption level leads to significant growth in the wealth of country.


\begin{thebibliography}{10}
\bibitem{tratio} To test at the 0.05 sugnificance level  
 if exponent $\lambda$ obtained from 
the regression line is statistically significant, we use   
 t-ratio (t-value) defined as $t=\lambda / \sigma$, where $\sigma$ 
 represents the standard deviation of the coefficient $\lambda$. 
 If $ t $ lies  outside the interval $-t_{0.975}$ to $t_{0.975}$,   
 where $t_{0.975}$ is a critical value, then $\lambda$ is
  statistically significant ($\lambda \ne 0$).
\end{thebibliography}
\end{document}